\documentclass[namedreferences]{solarphysics}
\usepackage[optionalrh]{spr-sola-addons} % For Solar Physics
\usepackage{graphicx}
\usepackage{url}          % For breaking URLs easily trough lines
\newcommand{\aap}{    {\it Astron. Astrophys.}}

\newcommand{\apj}{    {\it Astrophys. J.}}
\newcommand{\apjl}{   {\it Astrophys. J. Lett.}}
\newcommand{\apss}{   {\it Astrophys. Spa. Sci.}}

\newcommand{\jgr}{    {\it J. Geophys. Res.}}

\newcommand{\pasj}{   {\it Publ. Astron. Soc. Japan}}

\newcommand{\solphys}{{\it Solar Phys.}}

\newcommand{\ssr}{    {\it Space Sci. Rev.}}

\begin{document}
\begin{article}
\begin{opening}

\title{The 26 December 2001 Solar Event Responsible for GLE63.
I.~Observations of a Major Long-Duration Flare with the Siberian
Solar Radio Telescope}

\author{V.V.~\surname{Grechnev}\sep
        A.A.~\surname{Kochanov}}

 \runningauthor{Grechnev and Kochanov}
 \runningtitle{SSRT observations of a major flare}

\institute{Institute of Solar-Terrestrial Physics SB RAS,
                  Lermontov St.\ 126A, Irkutsk 664033, Russia
                  email: \url{grechnev@iszf.irk.ru}
                  email: \url{kochanov@iszf.irk.ru}}

\date{Received ; accepted }

\begin{abstract}

Ground Level Enhancements (GLEs) of cosmic-ray intensity occur, on
average, once a year. Due to their rareness, studying the solar
sources of GLEs is especially important to approach understanding
their origin. The SOL2001-12-26 eruptive-flare event responsible
for GLE63 seems to be challenging in some aspects. Deficient
observations limited its understanding. Analysis of extra
observations found for this event provided new results shading
light on the flare. This article addresses the observations of
this flare with the \textit{Siberian Solar Radio Telescope}
(SSRT). Taking advantage of its instrumental characteristics, we
analyze the detailed SSRT observations of a major long-duration
flare at 5.7 GHz without cleaning the images. The analysis
confirms that the source of GLE63 was associated with an event in
active region 9742 that comprised two flares. The first flare
(04:30\,--\,05:03 UT) reached a GOES importance of about M1.6. Two
microwave sources were observed, whose brightness temperatures at
5.7 GHz exceeded 10 MK. The main flare, up to the M7.1 importance,
started at 05:04 UT, and occurred in strong magnetic fields. The
observed microwave sources reached about 250 MK. They were not
static. Having appeared on the weaker-field periphery of the
active region, the microwave sources moved toward each other
nearly along the magnetic neutral line, approaching a
stronger-field core of the active region, and then moved away from
the neutral line like expanding ribbons. These motions rule out an
association of the non-thermal microwave sources with a single
flaring loop.
\end{abstract}

\end{opening}

\section{Introduction}
\label{S-introduction}

Solar energetic particles are somehow accelerated in association
with solar eruptive events. The highest-energy particles sometimes
produce in the Earth's atmosphere considerable fluxes of secondary
particles, which are able to cause ground-level enhancements (GLE)
of cosmic-ray intensity. GLEs are mainly detected with neutron
monitors (see, \textit{e.g.}, \citealp{Cliver2006, Nitta2012}; and
references therein).

Seventy-two GLE events caused by relativistic solar protons have
been observed from 1942 to 2015. Most solar sources of the 51 GLEs
registered after 1970 \citep{Kurt2004, Cliver2006, Aschwanden2012,
Gopalswamy2012, Gopalswamy2013, Thakur2014} were associated with
major flares of the soft X-ray (SXR) GOES X class (40 GLEs,
including GLE72 -- see \citealp{Chertok2015}) or M class (GLE28 with
an M5, GLE63 with an M7.1, and GLE71 with an M5.1 flare). The GOES
importance of the flare associated with GLE24 and those of the
far-side sources of GLE23, GLE29, GLE39, GLE50, and GLE61 are
uncertain. Atypically favorable conditions probably accounted for
GLE33 (C6) and GLE35 (M1.3) associated with moderate flares, weak
microwave bursts, and relatively slow coronal mass ejections (CMEs),
due to shock-acceleration of a high coronal seed population
\citep{Cliver2006}.

The rare occurrence of GLEs and their frequent association with both
big flares and fast CMEs hampers identifying their origins and makes
studying parent solar events highly important. The `big flare
syndrome' concept \citep{Kahler1982} explained the correlation
between the parameters of near-Earth proton enhancements and flare
emission by a general correspondence between the energy release in
an eruptive flare and its various manifestations. However, recent
studies by \cite{Dierckxsens2015} and \cite{Trottet2015} indicate
that both flares and shock waves can accelerate GLE particles.
\cite{Grechnev2013b} revealed a scattered correlation between the
peak fluxes of $>100$~MeV protons and peak flux densities of 35~GHz
bursts, although the proton outcome of four events, including GLE63
and GLE71, was much stronger. Further, \cite{Grechnev2015} found a
higher correlation between the total proton and microwave fluences;
nevertheless, the proton-abundant events were the outliers. Their
superiority could be due to, for example, predominant
shock-acceleration or contributions from stronger nearly concurrent
far-side events.

The subject of the present study is the 26 December 2001 solar event
with an SXR peak time at 05:40~UT (all times hereafter refer to
UT) responsible for GLE63. The aim is to understand the possible
causes of its atypically high proton outcome. Some other aspects of
this solar event look also challenging.

The SXR emission of this event was atypically long
\citep{Aschwanden2012}. Examination of the SXR light curves led
\cite{Gopalswamy2012} to set a probable onset time of the associated
flare at 05:03~UT. This time is close to the extrapolated CME onset
time in the online CME catalog
(\url{http://cdaw.gsfc.nasa.gov/CME_list/}; \citealp{Yashiro2004})
and a reported appearance of a type II burst in the range
04:59\,--\,05:02~UT. On the other hand, this burst is clearly visible
at 04:53~UT and detectable still earlier in the \textit{Hiraiso Radio
Spectrograph} (HiRAS) spectrum (2001122605.gif) at
\url{http://sunbase.nict.go.jp/solar/denpa/hirasDB/Events/2001/}.
This slowly drifting burst evidences the presence of a moving source
at least, ten minutes before the estimated onset time of a fast CME
(average speed of 1446~km~s$^{-1}$ according to the CME catalog).
Furthermore, it is not clear why the fast CME and a strong shock
wave (possibly responsible for the GLE particles) developed in
association with a microwave burst, which was not extremely strong.
It is not also clear when and where the shock wave appeared.

The flare and eruption in this event have been studied incompletely
because of limited data. The observations with the
\textit{Extreme-ultraviolet Imaging Telescope} (EIT:
\citealp{Delaboudiniere1995}), onboard the \textit{Solar and
Heliospheric Observatory} (SOHO), had a gap from 04:47 to 05:22~UT. The
\textit{Transition Region and Coronal Explorer} (TRACE:
\citealp{Handy1999}) did not produce extreme-ultraviolet images in
which an eruption could be detected. No SXR images or hard X-ray
data are available.

The search for extra data revealed that the 26 December 2001 event
was observed in microwaves with the \textit{Siberian Solar Radio
Telescope} (SSRT: \citealp{Smolkov1986, Grechnev2003}) at 5.7~GHz;
the \textit{Nobeyama Radioheliograph} (NoRH; \citealp{Nakajima1994})
at 17 and 34 GHz, and in 1600~\AA\ by TRACE. The analysis of extra
observations found for this event has led to new results that we
present in three companion papers.

This article (Paper~I) addresses the SSRT observations of this
flare. The SSRT routinely observes the Sun since 1996, but the
difficulties to clean and calibrate the SSRT images of strong flare
sources and a rather long time (typically 2\,--\,3 min) required to
produce each image restrict the opportunities to use its imaging
(2D-mode) observations. The usage of the SSRT data in studies of
flares has been limited (\textit{e.g.} \citealp{Altyntsev2002,
Altyntsev2007, Altyntsev2016, Meshalkina2012, Alissandrakis2013}).
The 1D-mode observations of a higher time resolution have also been
involved, especially in studies of microwave sources on short time
scales.

The flare occurred near the winter solstice that is a most
unfavorable season for the observations with the SSRT. At that
time, its beam pattern is considerably extended in the north-south
direction, and distortions associated with insufficiently accurate
knowledge of instrumental characteristics reach maximum. These
circumstances hamper efficient cleaning of the images.
Nevertheless, a relatively low level of the SSRT beam side lobes
due to a large number of antennas arranged in an equidistant array
makes raw images (without cleaning) usable for an analysis.

In the present article the 26 December 2001 flare is studied from
SSRT raw images as well as its 1D response. This has made possible
to study the flare development in microwaves, to produce detailed
light curves of the total flux and brightness temperature of the
microwave sources at 5.7~GHz, and to analyze their spatial
evolution. We focus on the opportunities provided by the SSRT data
in studies of major flares and pursue the results, which can be
obtained from the SSRT observations using the simplest estimations.
Some of the results, which do not correspond to conventional
properties of microwave flare sources, will be analyzed using
observations in different spectral domains.

Analyzing the TRACE and NoRH observations along with multi-frequency
total flux data, Paper~II \citep{Grechnev2016a} elaborates and
clarifies the conclusions drawn in the present article. Based on the
results of these two articles, Paper~III (Grechnev \textit{et al.},
2016b, in preparation) will address the eruption and endeavor to
understand the possible causes of the high proton productivity of
the 26 December 2001 solar event.

\section{Observations}

\subsection{Summary of the Flare}
\label{S-Overview}

\begin{figure} % {4}
  \centerline{\includegraphics[width=\textwidth]
   {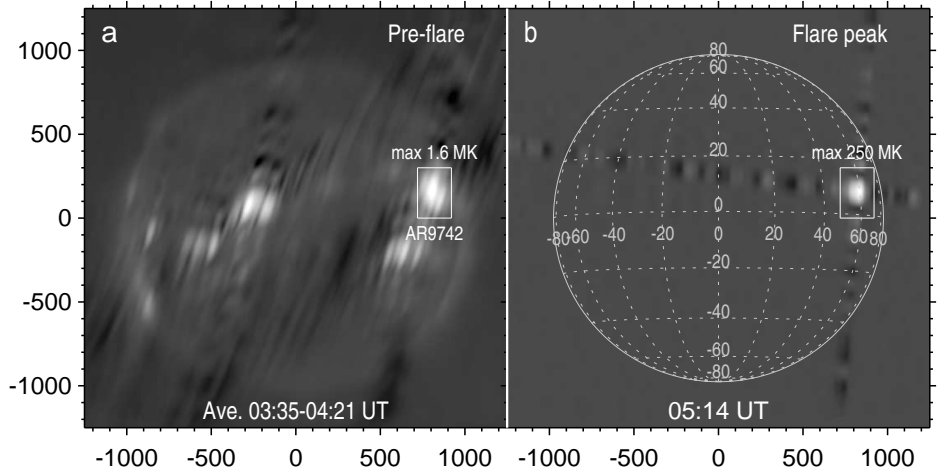}
  }
  \caption{SSRT full-disk images (logarithmic brightness scale).
(a)~Averaged pre-flare image. (b)~Single image at the flare peak.
The maximum brightness temperatures over AR~9742, which is shown
within a white rectangle, are specified in the panels. The axes
indicate arc seconds from the solar disk center.}
  \label{F-ssrt_fulldisk}
  \end{figure}

The event occurred in active region (AR) 9742 (N08\,W54). The AR had
a $\beta \gamma$ magnetic configuration.
Figure~\ref{F-ssrt_fulldisk}a shows an average of 31 raw microwave
images produced for the pre-flare interval of 03:35\,--\,04:21~UT using
a recent SSRT software \citep{Kochanov2013}. Despite instrumental
distortions, the average image shows both active regions and the
solar disk. A moderate level of stripes is due to relatively low
side lobes; the level of the first (negative) side lobes of a
perfectly phase-aligned SSRT antenna array is $-22\%$, of the second
ones is $+13\%$, \textit{etc.} The side lobes from an extended
source become smoothed and reduced.

The quiet-Sun brightness temperature at 5.7~GHz is $16 \times
10^3$~K. The brightness temperatures of the microwave sources in
AR~9742 (see the white rectangle) did not exceed 1.6~MK. A single
image in Figure~\ref{F-ssrt_fulldisk}b presents flare sources at the
peak of the microwave burst (05:14~UT). Their brightness temperatures
reached about 250~MK, so that all the other weaker sources
disappeared in the image. The stripes in this image are weaker than
in Figure~\ref{F-ssrt_fulldisk}a; they do not exceed 11\% in the
north-south direction and 7\% in the east-west direction. This fact
indicates rather large size of the sources.

\begin{figure} % {4}
  \centerline{\includegraphics[width=0.85\textwidth]
   {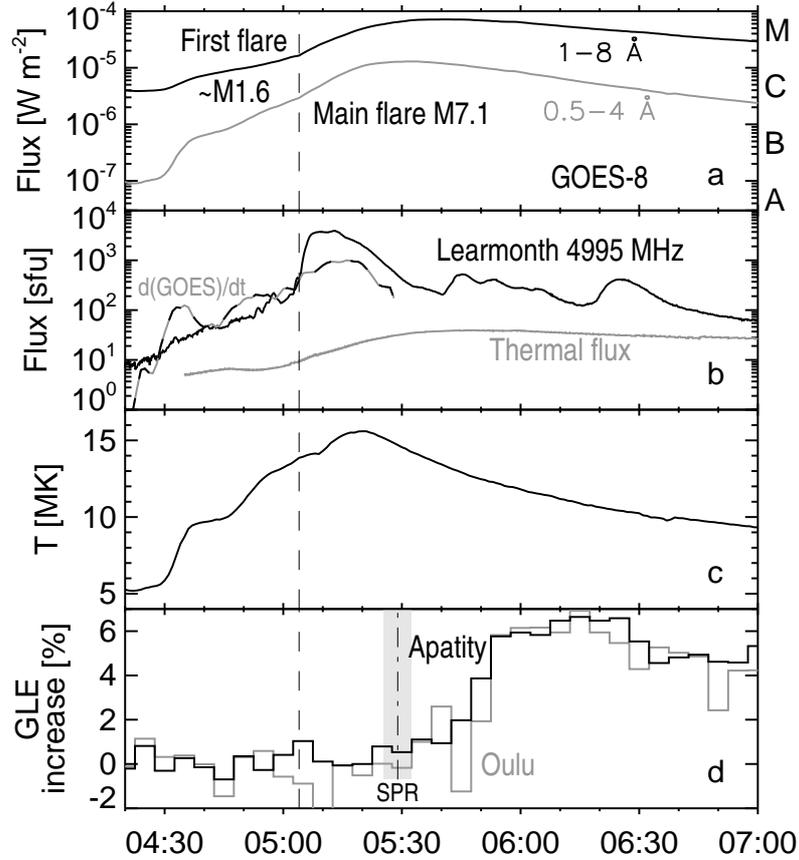}
  }
  \caption{Light curves of the event. (a)~SXR emission recorded by
GOES-8. (b)~Microwave burst. The gray curve represents the thermal
bremsstrahlung flux estimated from GOES data. The gray curve with
black dashes shows the time-derivative of the 0.5\,--\,4~\AA\ GOES
flux (positive part). (c)~Temperature computed from the GOES data.
(d)~GLE increase recorded by the Apatity (black) and Oulu (gray)
neutron monitors. The vertical dash-dotted line with shading denotes
the solar particle release time with uncertainties
\citep{Reames2009}. The vertical dashed line in all the panels
separates the first flare (up to about M1.6 importance) and the main
flare. All times refer to UT.}
  \label{F-flare_light_curves}
  \end{figure}

GOES-8 recorded a long-duration SXR emission starting at 04:32~UT that
reached up to M7.1 at 05:40~UT (Figure~\ref{F-flare_light_curves}a).
The temperature calculated from the two GOES channels is shown in
Figure~\ref{F-flare_light_curves}c. The calculated total emission
measure (not shown) reached $1.1 \times 10^{49}$~cm$^{-3}$ at 05:03~UT
(vertical dashed line) and a maximum of $4.6 \times
10^{49}$~cm$^{-3}$ at 05:47~UT.

A microwave burst in Figure~\ref{F-flare_light_curves}b was recorded
in Learmonth (US Air Force \textit{Radio Solar Telescope Network})
at a frequency of 4995~MHz, which is close to the observing
frequency range of the SSRT. The pre-burst level was subtracted. A
moderate increase of the microwave emission by 05:02~UT changed to a
sharp rise up to about 4000~sfu at 05:08\,--\,05:13~UT. The gray curve
represents thermal bremsstrahlung flux calculated from the
temperature and emission measure. The microwave burst was dominated
by non-thermal emission. The contribution from thermal
bremsstrahlung became significant late at the decay phase. The gray
curve with black dashes presents a positive part (by 05:28~UT) of the
time-derivative of the 0.5\,--\,4~\AA\ flux scaled to be comparable
with the microwave burst. The similarity between the shapes of the
GOES derivative and the microwave burst (called the Neupert effect;
\citealp{Neupert1968}) supports their common source. In summary, the
SXR and microwave emissions indicate that the electron acceleration
in the flare started by 04:30~UT and strengthened at 05:03~UT.

Figure~\ref{F-flare_light_curves}d presents the 5-min data from the
Apatity (black) and Oulu (gray) neutron monitors, which recorded
GL63 up to 6.5\%. The vertical dashed-dotted line marks the solar
particle release (SPR) time estimated by \cite{Reames2009} from the
velocity dispersion analysis. The shading indicates the SPR
uncertainty. The delay between the flare emission and the estimated
SPR time might indicate that acceleration of heavy particles was not
efficient enough before 05:30~UT, or that they were accelerated
simultaneously with electrons, but got an access to open magnetic
fields to reach Earth after that time only. Some observational
indications related to this issue will be considered in Paper~III.

As mentioned, the SXR emission of the flare was atypically long. Its
half-height duration in 1\,--\,8~\AA\ was 86~min, being twice longer
than the other longest GLE-related flares of Solar Cycles 23 and 24
(44~min for GLE62, 39~min for GLE71, 37~min for GLE69, 29~min for
GLE68 and GLE58; all others were still shorter). Both the rise and
decay phases of the flare had very long durations. Examining the
logarithmic time derivative of the SXR intensity,
\cite{Gopalswamy2012} estimated the onset time of the GLE-associated
flare at 05:03~UT (the vertical dashed line). From the presence of the
long preceding SXR emission up to about M1.6 importance and an inflection
in each of the GOES light curves at that time, \cite{Grechnev2013b}
supposed the possible occurrence of a nearly concurrent stronger
far-side event. It was difficult to verify this conjecture due to
the lack of images, in which eruptions could be detected. Most
likely, this conjecture is not confirmed; the microwave burst in
Figure~\ref{F-flare_light_curves}b evidences an increasing
non-thermal emission up to 200~sfu at 05:02~UT, corresponding to the
GOES derivative. If the only source of the microwave burst was
AR~9742, then the only source of the SXR emission was the flare in
this region. We will check this using the SSRT data.

\subsection{Raw SSRT Data}
\label{S-raw_data}

The SSRT is located at geographic coordinates $\phi = $~N$51^{\circ}
45^{\prime}$, $\lambda = $~E$102^{\circ} 13^{\prime}$. From 1996 to
July 2013, the SSRT operated in its initial design. At that period,
including 26 December 2001, the SSRT was a cross-shaped
interferometer consisting of two equidistant linear arrays, EW and
NS, each of 128 antennas spaced by $d = 4.9$~m. The observing
frequency band was from 5.67 to 5.79 GHz (central frequency
$\nu_\mathrm{C} = 5.73$ GHz). Unlike modern synthesis
interferometers, the SSRT was a directly-imaging telescope. Imaging
was performed using the frequency dependence of the SSRT beam
direction (frequency scanning). Each image of the Sun was formed in
its passage through the multi-beam fan due to the diurnal rotation
of the Earth \citep{Smolkov1986, Grechnev2003}.

The signals from all antennas of each linear array came into the
waveguide system, which had the structure of a binary tree, and were
combined to form NS and EW output signals. The 2D correlation
component was extracted using a well-known modulation technique. The
phases of the output signals of the NS and EW linear interferometers
were modulated and the signals were combined into in-phase and
anti-phase sums. The construction existing since late 1990s did not
allow extracting a response from each linear interferometer.

Due to the summation of all the antenna's signals \textit{in situ}
in the waveguide system, their dynamic range corresponded to the
actual range of brightness temperatures of microwave sources and
could be very high. For example, the dynamic range of the images in
Figure~\ref{F-ssrt_fulldisk} exceeds $2 \times 10^4$. The high input
dynamic range is a critical point of a directly imaging telescope,
thus requiring a controllable attenuator.

\begin{figure} % {4}
  \centerline{\includegraphics[width=\textwidth]
   {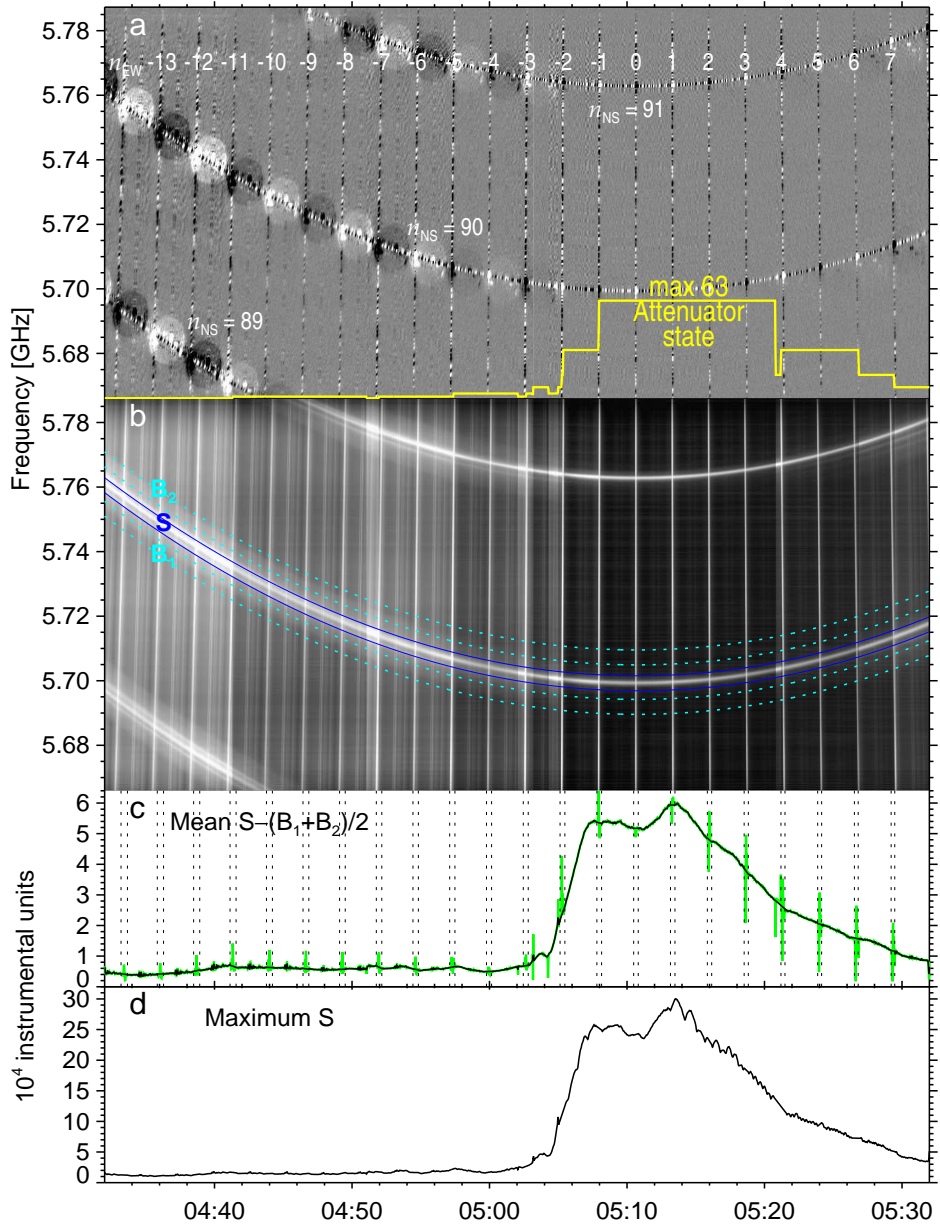}
  }
  \caption{Raw SSRT data. (a)~2D-mode data. The nearly vertical stripes
correspond to different interference orders of the EW interferometer
specified in the panel. Three long arcs correspond to interference
orders 89, 90, and 91 of the NS interferometer. The yellow plot in
panel (a) represents the attenuator state. (b)~1D-mode data. The
long arc-like band S within the dark-blue arcs traces the NS
response to the flaring source. The light-blue dotted bands B$_1$
and B$_2$ are used to compute the background. (c)~Variations of an
average over the S band with a subtracted averaged background.
(d)~Variations of the maximum over band S.}
  \label{F-ssrt_record}
  \end{figure}

The receiver system of the SSRT was a spectrum analyzer. Its output
presented the solar images in the time\,--\,frequency coordinates. A
portion of raw SSRT data on 26 December 2001 with a main part of the
microwave burst is shown in Figure~\ref{F-ssrt_record}a (the
difference of the in-phase and anti-phase signals) and in
Figure~\ref{F-ssrt_record}b (the sum of these signals). Each
frequency (total 500 channels) corresponds to a different viewing
direction. The sampling interval was 0.336~s in routine
observations. To follow the development of the burst,
Figure~\ref{F-ssrt_record}d presents the variations of the maximum
brightness along the dashed arc in Figure~\ref{F-ssrt_record}b. A
way to compute this time profile is discussed later in this section.

The beam of a linear interferometer depends on a direction cosine
[$\cos \theta $] with $\theta$ being a viewing direction relative to
the interferometer's baseline. For the EW interferometer, $\cos
\theta_\mathrm{EW} = \sin H \cos \delta$; for the NS interferometer,
$\cos \theta_\mathrm{NS} = \cos H \cos \delta \sin \phi - \sin
\delta \cos \phi$, with $H$ being the hour angle and $\delta$ the
declination. Each linear interferometer has major maxima in the
directions where $\cos \theta _n = nc/(\nu d)$, with numbers
$n=0,\pm 1,\ldots ,\pm n_{\max }$, $c$ the speed of light, $\nu$ a
frequency, and $n$ the interference order ($n_{\max}=\nu d /c =
93$).

The portion of the SSRT observation in Figure~\ref{F-ssrt_record}
contains the local noon (near 05:12~UT). At this time $H = 0$,
$n_\mathrm{EW} = 0$, the frequency dependence of the EW
interferometer's spatial sensitivity disappears and its 1D response
to a compact source in the time\,--\,frequency coordinates is a
vertical stripe. The 1D EW responses in the other interference
orders, from $-14$ to $+7$, appear as a set of nearly vertical
stripes. The 1D response of the NS interferometer appears as long
arcs, symmetric relative to the local noon. Figures
\ref{F-ssrt_record}a and \ref{F-ssrt_record}b contain orders
$n_\mathrm{NS} = $ 89, 90, and 91.

The 2D response in Figure~\ref{F-ssrt_record}a is formed in the
passage of the Sun through successive interference orders (for each
linear interferometer), which have alternate signs. The small dark
and light ellipses at intersections of the interference maxima of
the NS and EW interferometers are the corresponding images of the
Sun. The imaging software \citep{Kochanov2013} extracts each image
with surroundings corresponding to a required field of view and
transforms it to rectangular coordinates in the plane of the sky. As
the burst rose, the attenuator decreased the signal (yellow plot, 18
changes in this interval). The ellipses faded relative to the burst
source and disappeared right of $n_\mathrm{EW} = -3$, after 05:04~UT.
Faint non-flaring structures started to reappear right of
$n_\mathrm{EW} = 5$, after 05:24~UT.

The 1D response in Figure~\ref{F-ssrt_record}b contains the summed
additive signals from both interferometers, while the 2D correlation
component is suppressed. The basic interference structure of the 1D
data is the same as that of the 2D data. Here each sample at a given
frequency represents an instantaneous total of the solar emission
over a narrow stripe extracted by the EW interferometer's 1D-mode
knife-edge beam summed with similar total coming from the NS
interferometer.

A long bright arc in Figure~\ref{F-ssrt_record}b labeled S is a
trace of the flare region observed by the NS interferometer in
$n_\mathrm{NS} = 90$. Appropriate corrections were introduced to
compensate for the displacement of this source from the solar disk
center. Portions of similar traces correspond to $n_\mathrm{NS} =
89$ (before 04:44~UT) and $n_\mathrm{NS} = 91$ (after 04:45~UT). The
bright response in $n_\mathrm{NS} = 90$ within the dark-blue arcs is
a sum of the flaring microwave sources and a stripe of the quiet
solar disk within the one-dimensional NS interferometer's beam.

When the flare emission became strong enough, the contribution from
the quiet-Sun stripe can be neglected and the response from band S
within the dark-blue arcs reproduces approximately the total
microwave flux. A corresponding plot corrected for the attenuator
state changes is shown by the green line in
Figure~\ref{F-ssrt_record}c. The vertical broken lines delimit the
intervals, in which the responses to the flare region from the EW
interferometer appear. To reduce their contribution, a half-sum of
the background bands, $B_1$ and $B_2$ (of the same widths as band
S), is subtracted. Then the light curve was smoothed with a boxcar
of 11~s and interpolated within the intervals between the pairs of
the vertical broken lines. The result is shown with the black curve.

Similarly, Figure~\ref{F-ssrt_record}d shows the variations of an
instant maximum over the S band. As long as the brightness
distribution among the microwave flare sources is nearly constant,
this plot reproduces the temporal variations of their maximum
brightness temperature. A modulation with a decreasing period, that
is especially clearly visible after 05:13~UT, is possibly caused by the
passage of the NS interferometer's response across the sensitive
cells of the spectrum analyzer. A gradual shape of this curve
justifies the interpolation applied to the total-flux plot in
Figure~\ref{F-ssrt_record}c. The similarity of the two plots during
the major burst indicates the constancy of the total area of the
microwave sources. The initial parts of the plots, where the
quiet-Sun contribution is significant, are less similar.

\subsection{Calibration}
\label{S-calibration}

The SSRT images are calibrated referring to the quiet-Sun brightness
temperature of $T_\mathrm{QS} =$~16\,000~K \citep{Zirin1991,
Borovik1994}. The calibration technique is based on the analysis of
the brightness distribution in an image, where two statistical peaks
should be present. One of them corresponds to the zero sky level and
the second corresponds to the quiet-Sun level. The images are
calibrated by referring to the positions of the maximum values in
the two peaks (see \citealp{Kochanov2013} for details). For brevity,
henceforth we refer to this technique as auto-calibration. It works,
as long as the microwave sources are not extremely bright, so that
the quiet-Sun disk is detectable.

We have produced 105 SSRT images starting from the pre-flare stage
at 03:35~UT (Figure~\ref{F-ssrt_fulldisk}a) up to the late decay, by
06:28~UT. Most of them were auto-calibrated, excluding the major part
of the burst. To complement this interval with calibrated SSRT
data, we invoke the time profiles computed from 1D data shown in
Figures \ref{F-ssrt_record}c and \ref{F-ssrt_record}d. These light
curves are quantified in instrumental units. It is possible to
express the time profile of the maximum brightness in absolute
brightness temperatures by referring to the 2D images, using the
rise phase of the burst, when both kinds of data are available.

\begin{figure} % {2}
  \centerline{\includegraphics[width=0.7\textwidth]
   {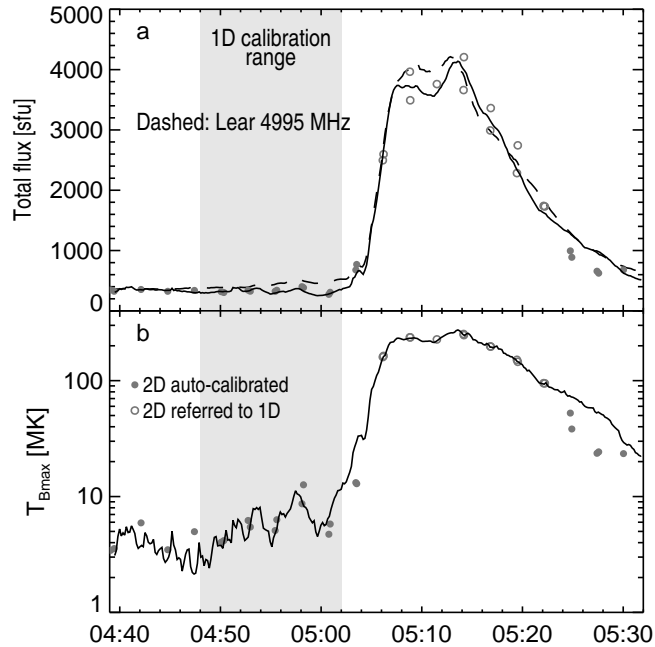}
  }
  \caption{Time profiles of the microwave emission at 5.7 GHz produced
from SSRT data: (a)~total flux and (b)~maximum brightness
temperature. The dashed curve in panel (a) represents the Learmonth
data at 4995 MHz for comparison. The circles represent the
measurements from imaging SSRT data: filled gray circles correspond
to auto-calibrated images and open circles represent the images
calibrated by referring to 1D data. The shading denotes the interval
in which 1D data were calibrated by referring to 2D images.}
  \label{F-ssrt_timeprof}
  \end{figure}

Figure~\ref{F-ssrt_timeprof}b illustrates the procedure. The
maximum brightness temperatures found from auto-calibrated images
are shown by the filled circles. The shading indicates the
calibration interval of the 1D data. Two images around 05:03~UT were
degraded by the attenuator shifts and excluded from the
calibration procedure. The calibration coefficients were found by
means of a linear regression. Using them, we have obtained the
calibrated time profile of the maximum brightness temperature
shown by the solid line. Its initial variations might not be
reliable due to contributions from the other sources elsewhere on
the Sun. By referring to this time profile, we calibrated the 2D
images in brightness temperatures during the major burst (open
circles).

When all of the 2D SSRT images have been calibrated, the total flux
time-profile can be quantified in sfu. This is shown in
Figure~\ref{F-ssrt_timeprof}a by the filled circles for
auto-calibrated images and by the open circles for the images
calibrated by referring to the 1D SSRT data. A relatively wide
scatter is mostly due to shortcomings of the imaging software. Some
of them were not revealed previously because of deficient experience
in handling major flares. Some distortions are due to insufficiently
accurate knowledge of characteristics of the SSRT systems. They are
strongest when observing the Sun at low altitudes, which was the
case on 26 December 2011. These causes and multiple changes of the
attenuator state probably resulted in the deviations of three of the
latest images.

At a final step, we have calibrated the total flux time-profile
computed from 1D SSRT data by referring to the 2D images in the
whole interval and using the same regression technique (solid line
in Figure~\ref{F-ssrt_timeprof}a). To evaluate the calibration
quality, the dashed line shows the total flux recorded in Learmonth
at 4995 MHz. The turnover frequency of the microwave burst was close
to the SSRT observation range (\citealp{Grechnev2013b}; Paper~II);
therefore, the total fluxes recorded by the SSRT at 5.7 GHz and in
Learmonth at 4995 MHz should almost coincide. Actually, with all
inaccuracies of the measurements from the 1D SSRT data, the
difference between the time profiles produced from them and the
Learmonth data is within 15\%. The scatter of the 2D SSRT data,
excluding the latest problematic images, is within 20\% (the
previous image-to-image calibration stability reached, at best,
30\%, \citealp{Grechnev2003}).

The brightness temperatures at 5.7 GHz exceeded 100~MK during the
major burst and reached 270~MK at the maximum, stronger than the
first flare by more than one order of magnitude. Adopting an
uncertainty range of $\pm 15\%$, \textit{i.e.} from 230~MK to
310~MK, we decided to choose a round value of 250~MK (lower than
the estimated 270~MK) within this range as a probable estimate of
a highest brightness temperature reached during the burst. Such a
high brightness temperature, never observed with the SSRT
previously, confirms the non-thermal nature of the emission
(\textit{cf.} Figure~\ref{F-flare_light_curves}c).

Observations of major flares with the SSRT were problematic
previously. It has become possible to produce the detailed microwave
time profiles in this event just due to the very strong emission,
which allowed neglecting a contribution from the other sources
elsewhere on the Sun. Conversely, this contribution makes the
approach applied here problematic in studies of weaker events.

\subsection{Microwave Sources}
\label{S-sources}

\subsubsection{Main Properties}
 \label{S-main_properties}

\begin{figure} % {1}
  \centerline{\includegraphics[width=\textwidth]
   {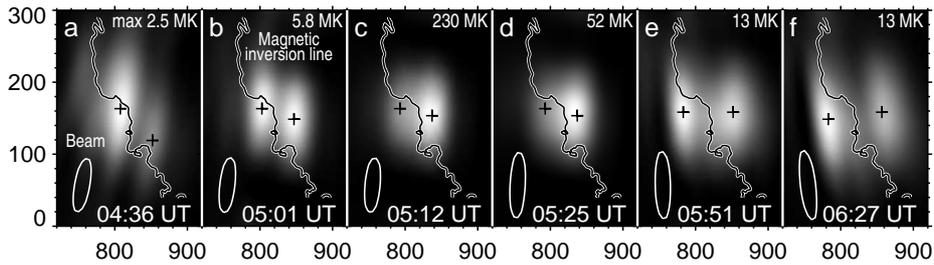}
   }
  \caption{Microwave images of AR~9742 at 5.7 GHz (linear brightness scale)
observed at different stages of the event: (a)~at the early onset
of the first flare; (b)~at its end, short before the major flare;
(c)~near the peak of the burst; (d)~at the decay of the major
burst; (e) and (f) during the late decay phase. The observation
times of the images are marked by the short bars in
Figure~\ref{F-2_sources}c. The field of view corresponds to the
white frame in Figure~\ref{F-ssrt_fulldisk}. The white ellipses
represent the half-height contours of the SSRT beam. The
black-on-white contour represents the magnetic polarity inversion
line. The crosses mark the brightness centers of the two sources.}
  \label{F-ssrt_img}
  \end{figure}

Figure~\ref{F-ssrt_img} shows selected microwave images of AR~9742
observed by the SSRT at 5.7~GHz at different stages of the event.
The field of view corresponds to the white frame in
Figure~\ref{F-ssrt_fulldisk}. The maximum brightness temperatures
over the images are specified in the panels. The white elliptic
contour presents the half-height SSRT beam. The black-on-white
contour traces the major magnetic polarity inversion (neutral) line
at the photospheric level separating an S-polarity east region from
an N-polarity west region. The neutral line was found from a
longitudinal magnetogram obtained with the \textit{Michelson Doppler
Imager} (MDI: \citealp{Scherrer1995}), onboard SOHO, on 26 December
at 04:51~UT.

All SSRT images show two distinct microwave sources. Their spatial
structures are unresolved; thus, the real brightness temperatures
were still higher. We strove to improve the coalignment accuracy
between SSRT images and their pointing. We found the positions of
the 5.7~GHz sources relative to the neutral line, comparing them
with properly processed higher-resolution NoRH 17~GHz images, TRACE
1600~\AA\ images, and white-light images from TRACE and SOHO/MDI.
The final pointing accuracy should be within $20^{\prime \prime}$.
The crosses in Figure~\ref{F-ssrt_img} mark the brightness centers
of the sources found by an automatic algorithm described in the next
section.

Figure~\ref{F-ssrt_img}a presents the onset of the first flare.
The major source here is close to the neutral line, polarized by
about $-7\%$, and does not have a counterpart at 17 GHz. These
properties are similar to those of the sources associated to the
neutral line, which are typically related to the origin of strong
flares (see, \textit{e.g.}, \citealp{Uralov2008}). A somewhat
increased brightness temperature of this source of 2.5~MK can be
due to an additional gyrosynchrotron emission from the flare,
which had already started. Another, weaker source (1.4~MK) is
located west of the neutral line. A faint extended feature in the
upper-left corner is due to a beam side lobe.

The centers of the sources in the other images were located at
different sides of the neutral line. Both sources were polarized. In
Figures \ref{F-ssrt_img}a\,--\,\ref{F-ssrt_img}c, the centers of the
sources move toward each other along the neutral line. The distance
between them decreases. Around the flare peak, the sources
overlapped (Figures \ref{F-ssrt_img}c and \ref{F-ssrt_img}d), so
that the distance between their centers could be somewhat
underestimated at that time. Then, the centers moved away from the
neutral line in Figures \ref{F-ssrt_img}d\,--\,\ref{F-ssrt_img}f.
The angle between the line connecting the centers of the sources and
the neutral line increased. A weaker diffuse brightening appeared
between the major sources in the late decay phase (Figures
\ref{F-ssrt_img}e and \ref{F-ssrt_img}f), possibly due to thermal
bremsstrahlung from the upper part of the flare arcade.

The relation between the brightness temperatures of the two
sources was not constant. The source associated to the neutral
line was brighter before the flare. The ratio of the maximum
brightness temperatures of the west to east source initially was
in the range 0.5\,--\,0.6, gradually increased to 1.0\,--\,1.1
during the first flare, reached 1.5\,--\,1.7 during the main
flare, and then decreased to 0.7\,--\,1.1. While the maximum
brightness temperatures changed during the event by two orders of
magnitude, a narrow range of the variations in the ratio between
the two sources indicates a similarity of their time profiles.

Comparison of the images in Figure~\ref{F-ssrt_img} with half-height
contours of the SSRT beam indicates that the sources were not
compact. Decomposition of the east\,--\,west and north\,--\,south
cross-sections of the image observed near the peak of the burst in
Figure~\ref{F-ssrt_img}c has resulted in an observed size of the
sources of $\approx 43^{\prime \prime} \times 94^{\prime \prime}$,
and a deconvolved size of $\approx 38^{\prime \prime} \times
59^{\prime \prime}$ (with a beam of $21^{\prime \prime} \times
73^{\prime \prime}$ at that time). A maximum brightness temperature
of the deconvolved strongest west source at that time should
approximately be proportional to the ratio between the observed area
and the deconvolved one, \textit{i.e.}, $230 \times 1.82 \approx
420$~MK.

\subsubsection{Motions of the Microwave Sources}

To measure the distances between the observed sources and their
orientation relative to the neutral line, we used a model elliptic
source of $43^{\prime \prime} \times 60^{\prime \prime}$ extended in
the north\,--\,south direction. The convolution of this source with
an idealized SSRT beam (the real beam is not known precisely) looks
more or less similar to the observed sources.

The algorithm measuring the positions of the sources operated in the
following way. At the first step, a starting estimate was found for
the position of the maximum over each image. A model response
(computed as the convolution of the model source with the SSRT beam)
was scaled to the maximum, placed into this position, and subtracted
from the image. The position of the brightness center was calculated
for the second source. Then, a model response corresponding to the
second source was subtracted and a final estimate for the position
of the first source was calculated. This procedure supplied the
arrays of the $(x, y)$ coordinates for the two sources measured from
105 images. To reduce the scatter in the measurements, both the $x$
and $y$ arrays of each source were filtered using the median over
three neighbors and smoothed with a boxcar average over five
neighbors. The parameters of the model source somehow affect the
results, but do not change them significantly.

\begin{figure} % {2}
  \centerline{\includegraphics[width=0.7\textwidth]
   {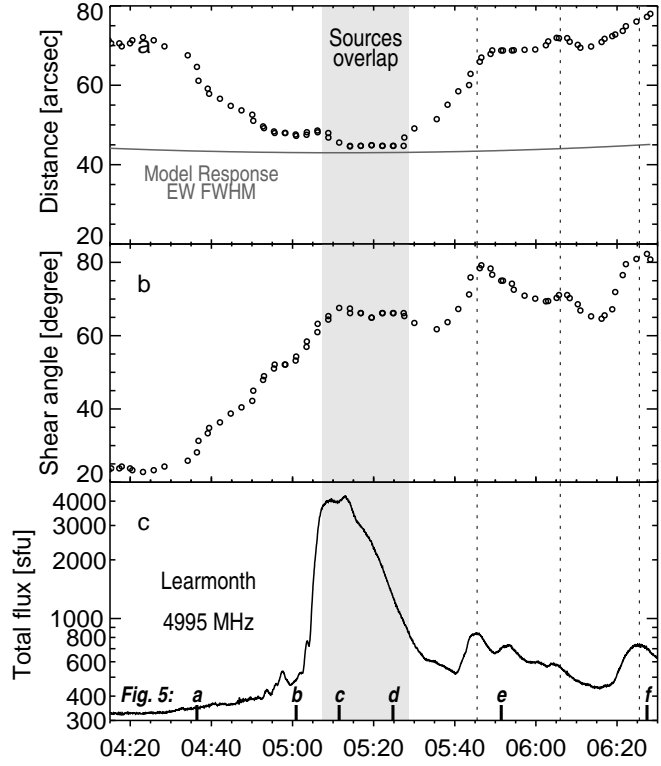}
  }
  \caption{Relative motions of the microwave sources
observed at 5.7 GHz. (a)~The distance between the brightness centers
of the two sources. The gray line represents a half-height width of
a model response in the east\,--\,west direction. The shading marks
the interval when the two sources overlap in the images. (b)~The
shear angle between the line connecting the brightness centers of
the two sources and the magnetic neutral line. (c)~The total flux
time-profile of the burst recorded at 4995 MHz in Learmonth
(including pre-burst emission). The vertical dotted lines mark some
minor features in the time profile. The short thick bars with labels
at the bottom denote the observation times of the images in the
corresponding panels of Figure~\ref{F-ssrt_img}.}
  \label{F-2_sources}
  \end{figure}

Figures \ref{F-2_sources}a and \ref{F-2_sources}b present the
detailed measurements of the relative positions and shear angle of
the two sources in a wide time interval. The angle is referred to a
major orientation of the neutral line of $105^{\circ}$ from the west
direction, without considering its detailed curves. The relative
measurements do not depend on the accuracy, with which the images
were coaligned with each other and with the magnetogram. To relate
the motions of the sources to the development of the flare,
Figure~\ref{F-2_sources}c presents the total flux time-profile of
the burst at 4995~MHz recorded at Learmonth. This time profile shows
various features in detail without a scatter, which is present in
the time profiles computed from the 2D SSRT data.

Along with the measured distances between the brightness centers of
the two sources, Figure~\ref{F-2_sources}a presents the full width
at half-maximum (FWHM) of a model response in the east\,--\,west
direction, where it was the narrowest. The half-height width of a
response puts the limit to resolve two features. If the two sources
are located too close, then the results can be biased. The
measurements within the shaded interval, when the sources overlap in
the images, should therefore be considered with caution.

The measured positions seem to be affected by subsidiary bursts,
whose sources might be located away from the brightness centers of
the major burst. These minor features in the time profile in
Figure~\ref{F-2_sources}c marked with the vertical dotted lines
have detectable counterparts in Figure~\ref{F-2_sources}b, and
some in Figure~\ref{F-2_sources}a.

The overall evolution corresponds to what can be deduced from the
images in Figure~\ref{F-ssrt_img}. The distance between the sources
in Figure~\ref{F-2_sources}a gradually decreased, starting at the
onset of the first flare, reached a minimum during the major burst,
and then gradually increased. The shear angle between the positions
of the microwave sources and the neutral line in
Figure~\ref{F-2_sources}b increased, starting at the onset of the
first flare, and after the major burst became non-monotonic, most
likely, due to subsidiary bursts.

\subsubsection{Degree of Polarization}

\begin{figure} % {2}
  \centerline{\includegraphics[width=0.7\textwidth]
   {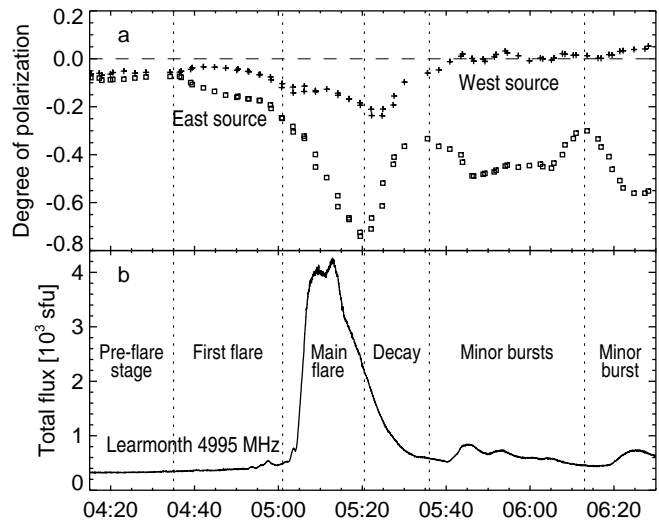}
  }
  \caption{Variations of the degree of polarization in the central
regions of the east (squares) and west (crosses) microwave sources
(a) in comparison with a total flux time profile (b). The vertical
dotted lines separate different episodes of the event indicated in
panel (b).}
  \label{F-polariz}
  \end{figure}

Despite the complexity of the polarization structure, the measured
positions of the brightness centers for the two sources allow
studying how the degree of polarization evolved. Its value was
measured for each source as an average within a small area
surrounding its brightness center. This area was defined by a
contour of the SSRT beam at a level of 0.65. The results are
practically insensitive to the changes of the level from 0.5 to 0.8,
which we tested. The measurements are related to the whole observed
configuration without subtraction of the quasi-stationary
background. To reduce the scatter, the data points were smoothed
with a boxcar average over five neighbors. The degree of
polarization calculated in this way does not depend on the accuracy
of the coalignment or calibration. The contributions from side
lobes, edge effects, and other instrumental issues are minimized
thereby. The results are presented in Figure~\ref{F-polariz}a in
comparison with the total flux time-profile in
Figure~\ref{F-polariz}b.

The time profiles in Figure~\ref{F-polariz}a show complex variations
of the polarization with a general correspondence to the main stages
of the event. This correspondence along with a gradual character of
the variations supports the reliability of the measurements from
each image. The parameters of the model source used to measure the
positions of the brightness centers somehow affect the results, but
do not change them significantly.

Before the flare onset, the degree of polarization was around $-7\%$
in both sources. The polarization of the east source started to
gradually strengthen with the onset of the first flare, reached
$-23\%$ at its end, and, during the main flare, strengthened
steeply. During the peak of the burst it increased from $-45\%$ to
$-60\%$, and reached $-75\%$, when the flux of the burst decreased
to a half-height level. During the subsequent decay, the
polarization of the east source weakened. Its later variations were
apparently related to subsidiary bursts.

The polarization of the west source was inverted and was
considerably weaker. Its variations from 04:50 to 05:36~UT were similar
to those of the east source. The maximum degree of polarization was
about $-20\%$ soon after the strongest polarization of the east
source. During the minor bursts after 05:36~UT, the variations of the
polarization in the two sources became dissimilar.

\section{Discussion and Summary}
 \label{S-discussion}

Both SXR and microwave data show that the long-duration flare in
AR~9742 consisted of two parts. The first flare started at about
04:30~UT, lasted half an hour, and reached a GOES importance of
about M1.6 at 05:03~UT. The main flare, which was much stronger in
microwaves, started at 05:04~UT and reached an importance of M7.1
at 05:40~UT. The major microwave burst fully developed and mainly
finished within 36~min, before the SXR peak. The decay of the SXR
emission possibly lengthened due to subsidiary bursts during
05:40\,--\,06:10~UT and around 06:25~UT. The SXR emission
corresponded to the microwave burst via the Neupert effect. The
only source of the microwave burst was AR~9742, and the only
source of the SXR emission was the flare in this region. Most
likely, the solar event in AR~9742 associated with the M7.1 flare
was the only source of the near-Earth proton event and GLE63.

In the preceding studies of microwave flare emissions, the sources
observed by the SSRT at 5.7~GHz were typically associated with
loop-top regions (\textit{e.g.} \citealp{Altyntsev2002,
Altyntsev2007, Altyntsev2016, Meshalkina2012}, and others). In the
26 December 2001 flare, two distinct microwave sources were
observed. They were polarized, comparable in size and brightness
temperatures (up to extremely high values), and varied rather
similarly. These properties of the two sources and their positions,
most likely at different sides of the magnetic neutral line,
indicate their association to the conjugate legs of a closed
magnetic structure.

Observations of flares in thermal emissions (H$\alpha$,
ultraviolet, extreme ultraviolet, soft X-rays, \textit{etc.})
typically show complex multi-loop structures. By contrast,
microwave and hard X-ray (HXR) images of non-thermal emissions in
impulsive flares usually reveal simpler configurations identified
with one or two loops (\textit{e.g.}, \citealp{Hanaoka1996,
Hanaoka1997, Nishio1997, GrechnevNakajima2002}; and many others).
Moreover, an analysis of the microwave morphologies in many
near-the-limb events led \cite{TzatzakisNindosAlissandrakis2008}
to a conclusion about single-loop microwave configurations
existing even in some long-duration major flares. These
observational indications resulted in a prevailing concept of a
single microwave-emitting flare loop (or, at most, two loops).
However, even the inhomogeneous flare-loop model, initially
proposed by \cite{Alissandrakis1984} and further developed using
powerful modeling tools to account for several inhomogeneities
\citep{TzatzakisNindosAlissandrakis2008,
KuznetsovNitaFleishman2011}, cannot explain various observations.
In particular, \cite{Zimovets2013} demonstrated that, at least,
some of the seemingly single loops shown by microwave NoRH images
corresponded to multi-loop arcades observed with telescopes of a
higher spatial resolution.

The concept of a single microwave-emitting flare loop is difficult
to reconcile with systematic motions of the microwave sources in
Figure~\ref{F-2_sources}a. A continuous relative displacement of the
two sources by $\gsim 20^{\prime \prime}$ in the plane of the sky
implies their association with different loops or loop systems at
different times. In principle, the observed relative motions of the
sources could be caused by a varying height of the emitting regions
in two legs of a single loop. This effect should also be manifested
in the degree of polarization due to a varying magnetic field
strength and viewing angle. However, while the variations in the
relative distance in Figure~\ref{F-2_sources}a before 05:20~UT are
nearly symmetric to those after 05:20~UT, the shear angle in
Figure~\ref{F-2_sources}b and polarization in
Figure~\ref{F-polariz}a are strongly asymmetric. Thus, while the
height variations are possible, the observed motions were most
likely determined by the displacements of the footpoint regions of
numerous arcade loops.

Similar motions are known from HXR observations. From a statistical
study of HXR sources, \cite{Bogachev2005} found different types of
their motions relative to the magnetic neutral line and interpreted
them in terms of the standard flare model.

The diverging motion at the decay phase (05:30\,--\,05:45~UT and later)
resembles a usual expansion of the flare ribbons, which represent
the footpoint regions of numerous loops. This is type~I motion
according to \cite{Bogachev2005}. The approach of the two sources
toward each other along the neutral line throughout the first flare
is also difficult to relate to a single loop; this kind of motion
was also observed in HXR (type~II motion in \citealp{Bogachev2005}).
The deviations from the systematic motions during subsidiary bursts
also imply more complex configurations than a single
microwave-emitting flare loop.

Additional indications are provided by the degree of polarization.
This parameter of the gyrosynchrotron emission is closely related to
the magnetic field strength, being not directly dependent on an
unknown number of emitting electrons. Keeping this in mind, the
varying degree of polarization in Figure~\ref{F-polariz}a is
difficult to understand in terms of the single-loop hypothesis. On
the other hand, with the limited data we consider, a probable
participation of additional loops makes the interpretation of these
complex variations ambiguous. We therefore consider, for simplicity,
the approximation of a single homogeneous source for each of the two
observed microwave sources.

To analyze the gyrosynchrotron emission, analytic approximations by
\cite{DulkMarsh1982} and \cite{Dulk1985} are widely used. The
accuracy of the formulas is reduced at low harmonics of the
gyrofrequency; in such situations, we invoke them to obtain rough
estimates and to understand major tendencies, and additionally refer
to the plots in those articles that present the results of numerical
calculations. Note that gyroresonance features in the spectrum are
not expected in observations due to inhomogeneities of the magnetic
field.

The polarization of the east source corresponded to the $x$-mode
emission and reached $r_\mathrm{c} \approx -0.75$. This is only
possible, if the optical thickness [$\tau$] was small ($\tau < 1$).
According to \cite{Dulk1985}, the degree of circular polarization,
$r_\mathrm{c}$, in the optically thin limit is
\begin{eqnarray}
r_\mathrm{c} \approx 1.26\times 10^{0.035\delta}\times
10^{-0.071\cos\theta}\left(\frac{\nu}{\nu_{B}}\right)^{-0.782+0.545\cos\theta}
\quad (\tau_{\nu}\ll 1), \nonumber
\end{eqnarray}
with $\delta$ being a power-law index of the number density spectrum
of microwave-emitting electrons, $\theta$ a viewing angle, and
$\nu_{B} \approx 2.8 \times 10^6 B$ the electron gyrofrequency in
the magnetic field [$B$]. The degree of polarization directly
depends on the magnetic field strength.

The polarization of the east source strengthened until the decay
phase. During the first flare, this occurred presumably due to an
increasing contribution of the gyrosynchrotron emission, which
became dominant at the end of the first flare
(Figure~\ref{F-flare_light_curves}b); and, possibly, due to the
motion of the source from a weaker-field periphery of the active
region to its stronger-field core. With a degree of polarization of
$-(0.20-0.25)$ just before the main flare and a viewing angle around
$\theta \approx 60^{\circ}$, corresponding to the position of
AR~9742, the magnetic field strength in the east source should not
exceed $-100$~G. This conclusion applies to the whole first flare.

An increasing degree of the polarization during the major burst
suggests a strengthened magnetic field in the east source. Assuming
for certainty $\theta = 60^{\circ}$ and $\delta = 2.5-3.5$ (see
Paper~II), with $r_\mathrm{c} \approx -0.5$ at the peak of the
burst, we estimate the magnetic field strength in the east source to
be around $-250$~G at that time. With $r_\mathrm{c} \approx -0.75$
at the end of the major burst, the magnetic field strength in the
coronal east source could reach $\approx -540$~G. This magnetic
field corresponds to $\nu/\nu_B \approx 4$, beyond the validity
range of the \cite{Dulk1985} approximation, but nevertheless
consistent with his Figure~3. If the optical thickness of this
source was not small enough to satisfy the condition $\tau_{5.7} \ll
1$, then the magnetic field should be somewhat stronger.

The magnetic field at the photosphere underneath should be
considerably stronger than in the low corona. The magnetic field in
this region of the MDI magnetogram on that day at 04:51~UT ranged from
$-600$ to $-850$~G. Since AR~9742 was not far from the limb, this
magnetogram might be strongly affected by the projection effect. We
additionally examined an MDI magnetogram observed two days before,
at 04:51~UT on 24 December. The magnetic fields at about this place
reached more than $-1000$~G; on the other hand, the active region
evolved. Thus, the estimated maximum magnetic field strength of
$\approx -540$~G in the east coronal source seems to be plausible.

The west source was polarized in the sense of the $o$-mode emission,
with a degree, not exceeding $-20\%$. Either its intrinsic emission
corresponded to the $x$-mode and was inverted, propagating through a
layer of the quasi-transversal magnetic field, or it was initially
optically thick. If the SSRT observing frequency, 5.7~GHz, was
higher than the peak frequency of this source, then its brightness
temperature should depend on the magnetic field strength directly
\citep{DulkMarsh1982, Dulk1985}. However, a weak magnetic field of
$-(40-45)$~G corresponding to $r_\mathrm{c} = -0.2$ with the same
$\delta = 2.5-3.5$ and $\theta = 60^{\circ}$ would contradict a
higher brightness temperature of the west source relative to the
east source. Hence, the west source was not optically thin. Its peak
frequency, $\nu_{\rm peak}$, was either slightly lower than 5.7~GHz
(inverted $x$-mode emission), or, most likely, higher (intrinsic
$o$-mode emission). The latter option is consistent with an
estimated $\nu_{\rm peak} \approx 6.9$~GHz for the total flux in
this event (\citealp{Grechnev2013b}; see also Paper~II). Indeed, the
total flux is the sum of the emissions from the east source with a
$\nu_{\rm peak} < 5.7$~GHz, and the west source, whose peak
frequency should be $> 5.7$~GHz, even if the total flux had
$\nu_{\rm peak} \geq 5.7$~GHz.

The peak frequency can be estimated, referring again to
\cite{DulkMarsh1982} and \cite{Dulk1985}, as
\begin{eqnarray}
\nu_\mathrm{peak} \approx 2.72 \times 10^3 \times
10^{0.27\delta}(\sin\theta)^{0.41+0.03\delta}(NL)^{0.32-0.03\delta}
\times B^{0.68+0.03\delta}, \nonumber
\end{eqnarray}
where $(NL)$ is a column density of emitting electrons. Although
it can be different in the two sources located in the conjugate
legs of the same closed structure, the dependence of
$\nu_\mathrm{peak} \propto (NL)^{0.22-0.25}$ is considerably
weaker than $\nu_\mathrm{peak} \propto B^{0.76-0.79}$. While the
basic formula might be inaccurate at a low harmonic of the
gyrofrequency, a stronger magnetic field seems nevertheless to be
a most probable reason for a higher $\nu_\mathrm{peak}$ in the
west source.

A brightness temperature of $\approx 4.2 \times 10^8$~K estimated in
Section~\ref{S-main_properties} for the deconvolved west source near
the peak of the burst (Figure~\ref{F-ssrt_img}c) roughly
corresponds, with $\delta = 2.5-3.5$ and $\theta = 60^{\circ}$, to
the optically thick emission around the fifth harmonic (Figures~3 in
\citealp{DulkMarsh1982} and \citealp{Dulk1985}), \textit{i.e.},
400~G \textit{vs.} $\approx 250$~G in the east source at the same
time. All of the estimates, along with a behavior of the
polarization during the main flare, indicate that the west source
was optically thick and located in a stronger, relative to the east
source, magnetic field, which increased in the course of the flare.
This conclusion is supported by a higher brightness of the west
source throughout the event at both 17 and 34~GHz in the NoRH movie,
\url{norh20011226_0505_pfi.mpeg}, available at
\url{http://solar.nro.nao.ac.jp/norh/html/event/} entry
\url{20011226_0505}. Both sources were optically thin at these two
frequencies, and therefore their brightness temperatures directly
depended on the magnetic field strength.

The magnetogram observed on 26 December shows the photospheric
magnetic fields in the west part of AR~9742 to be around 1000~G. A
sunspot was there. A nearly radial magnetic field in its central
part should be substantially reduced in the line-of-sight
magnetogram observed close to the limb. We have not radialized the
magnetogram to avoid overestimating the magnetic field in the region
under the east source. The MDI magnetogram on 24 December shows the
magnetic fields exceeding 2700~G in the central part of the sunspot.
Strong magnetic fields were really present on the photosphere
approximately under the west source, which was the brightest during
the main flare.

Some of the estimated magnetic field strengths fall outside the
range where the accuracy of the formulas by \cite{DulkMarsh1982} and
\cite{Dulk1985} is guaranteed. Nevertheless, our results are
supported by the following facts. i)~Our estimates are also
consistent with the results of numerical calculations by
\cite{DulkMarsh1982} and \cite{Dulk1985} in their Figures~3.
ii)~Comparison of the positions of the two sources in Figures
\ref{F-ssrt_img}b\,--\,\ref{F-ssrt_img}d with the magnetogram
confirms that magnetic fields under the west source were stronger
than those under the east source, as considered in this section,
while their probable values in the corona correspond to the
estimates. iii)~The NoRH movie of the flare observed at 17 and
34~GHz also supports our results.

We conclude that the main flare occurred in strong magnetic fields,
whose photospheric base, most likely, had a strength of $\gsim
1000$~G. Probably, the west flare ribbon extended into the strongest
magnetic fields above the sunspot. However, the spatial resolution
of the SSRT and the coalignment accuracy are insufficient to judge
to what extent this occurred. This issue will be addressed in
Paper~II, which will also analyze the microwave spectrum.

As mentioned in Section~\ref{S-introduction}, the onset time of the
main flare corresponds to an estimated launch time of the CME. The
major phase of the 26 December 2001 GLE63-related event resembles
those of the 20 January 2005 event (GLE69; \citealp{Grechnev2008})
and of the 13 December 2006 event (GLE70; \citealp{Grechnev2013a}).
Furthermore, \cite{Grechnev2013b} showed flaring in stronger
magnetic fields above the sunspot umbrae to be typical of big proton
events.

Like the previously mentioned events, the flare on 26 December 2001
involved rather strong magnetic fields and occurred, at least, close
to a sunspot. The magnetic fields involved in the GLE63-related
flare were probably not so strong as in the flares related to GLE69
and GLE70, when the peak frequencies exceeded 25 GHz, and the fluxes
at 35~GHz were considerably higher than $10^4$~sfu. Nevertheless,
major aspects of these events look qualitatively similar.

An additional particularity of the 26 December 2001 event was the
very long duration of the flare. The rise phase of the main flare
alone lasted 36~min \textit{vs.} 18~min for the flares related to
GLE69 and GLE70. According to the Neupert effect
\citep{Neupert1968}, this phase corresponds to the effective
particle acceleration in a flare. A considerably higher correlation
between the fluences of near-Earth proton enhancements, on the one
hand, and fluences of the SXR and microwave emissions, on the other
hand found by \cite{Grechnev2015} indicates a dependence of the
total number of high-energy protons arriving at the Earth orbit on
both the intensity and total duration of the acceleration process.
The role of the flare duration is obvious, if protons are
accelerated simultaneously with electrons in a flare, but it is more
difficult to expect such a correspondence, if protons are
accelerated by shock waves far away from a flare region. Therefore,
a considerably higher correlation between the fluences of protons
and flare emissions than between their peak values found for the 26
December 2001 event indicates a significant contribution from flare
processes to the acceleration of protons. The discussion of the
particle event will be further addressed in Paper~III.

\subsection{Conclusion}

Our analysis has confirmed that the solar event in active region
9742 associated with the M7.1 flare was the only source of the
near-Earth proton event and GLE63. No signs of a concurrent far-side
event have been found.

The event in AR~9742 consisted of two parts. The first flare
(04:30\,--\,05:03~UT) reached a GOES importance of about M1.6. The
brightness temperatures at 5.7~GHz exceeded 10~MK. The main flare,
up to M7.1 importance, started at 05:04~UT, when a CME was
launched. The microwave sources reached about 250~MK. The SSRT
data indicate that strong magnetic fields were involved in the
main flare. These magnetic fields were probably associated with
the sunspot in the west part of AR~9742.

The two microwave sources observed at 5.7~GHz initially approached
each other along the magnetic neutral line and then moved away from
it like expanding ribbons. These motions are difficult to understand
in the frame of a single-loop hypothesis. A natural explanation of
the observed properties of the microwave sources might be their
association to the legs of the flare arcade. To verify this
conjecture, microwave data should be compared with the flare arcade
or ribbons observed in a different spectral range, where they are
clearly visible. These issues will be addressed in Paper~II. The
possible causes of the high proton productivity of the 26 December
2001 event will be considered in Paper~III.

This is a first detailed study of a major long-duration flare from
combined 2D and 1D SSRT data. A relatively low side-lobe level of
the SSRT beam and rather large areas of the microwave sources
allowed using the images produced by the SSRT without cleaning. The
techniques described here provide an opportunity to study important
major flares recorded with the SSRT in the past. The analysis has
revealed shortcomings of the imaging and calibration software that
were not manifested previously because of a deficient experience in
handling major flares. Some imperfect techniques and software
(\textit{e.g.}, calibration routines) have been improved in the
course of our study. The development of some others is in progress.

The SSRT routinely carried out imaging observations of the whole Sun
at 5.7~GHz based on the initial operating principle from 1996 to
July 2013. Currently, the central part of the antenna array is under
reconstruction to upgrade the SSRT to the multi-frequency
(4\,--\,8~GHz) \textit{Siberian Radio Heliograph} (SRH:
\citealp{Lesovoi2012, Lesovoi2014}). The remaining part of the
original antenna array keeps on the initial-principle observations.

\begin{acks}

We thank A.T.~Altyntsev for the idea of this study and useful
remarks, and our colleagues for their contribution, efforts, and
assistance. The data used here are provided by the SSRT team in
Badary. S.A.~Anfinogentov has substantially contributed to the
collaborative development of the SSRT raw-data processing and
calibrating software and assisted in computations. S.V.~Lesovoy
developed the data acquisition system and the routine imaging
software. We thank him and A.M.~Uralov for fruitful discussions. We
appreciate the memories of T.A.~Treskov, one of the major developers
of the SSRT, whose ideas helped us to implement the techniques
described here, and N.N.~Kardapolova, who managed the SSRT
observations for many years. We thank the reviewer for useful
remarks.

We are grateful to the instrumental teams of SOHO/MDI (ESA and
NASA), GOES, USAF RSTN Network, and Nobeyama Radioheliograph.

A.K. was supported by the Russian Foundation of Basic Research
under grants 15-32-20504 mol-a-ved and 15-02-01089.

\end{acks}

\medskip

\noindent {\footnotesize \textbf{Disclosure of Potential Conflicts
of Interest} \quad The authors declare that they have no conflicts
of interest.}

\end{article}

\end{document}